\documentclass[aps,prl,twocolumn,floatfix]{revtex4}
\usepackage{graphicx,subfigure}
\usepackage{bm}

\begin{document}

\title{A Twisted Ladder: relating the Fe superconductors to the high $T_c$
cuprates}
\date{\today }

\begin{abstract}
We construct a 2-leg ladder model of an Fe-pnictide superconductor
and discuss its properties and relationship with the familiar
2-leg cuprate model. Our results suggest that the underlying
pairing mechanism for the Fe-pnictide superconductors is similar
to that for the cuprates.
\end{abstract}

\author{
E. Berg$^{1}$, S. A. Kivelson$^{1}$, and D.~J.~Scalapino$^{2}$ \\
$^1$Department of Physics, Stanford University, Stanford, CA
94305-4045, USA\\
$^2$Department of Physics, University of California, Santa
Barbara, CA 93106-9530, USA}

\maketitle

An important question has been raised by the discovery of high
temperature superconductivity (HTS) in the Fe-pnictides: Is there
a single general mechanism of HTS which operates, albeit with
material specific differences, in both the Fe-pnictides and the
cuprates (and possibly other novel superconductors), or do the
cuprates and the Fe-pnictides embody two of possibly many
different mechanisms of HTS? This question is complicated by one
of the perennial issue of the field: To what extent is it possible
to understand the properties of the strongly correlated electron
fluid in the Fe-pnictides, the cuprates, and other materials from
a weak or strong coupling perspective, given that the materials
exhibit some features that appear more natural in one limit and
some that are suggestive of the other. The Fe-pnictides, like
the cuprates, are likely in the intermediate coupling regime~\cite%
{Haule2008,Si1,Si2,Basov}, which is difficult to treat
theoretically.

In the present paper, we introduce a model of a two-leg ladder
with an electronic structure chosen to reproduce particular
momentum cuts through the Fe-pnictides band structure. We study
the magnetic and superconducting properties of this model
numerically using the density matrix renormalization group (DMRG)
\cite{White} method which allows us to treat the ``intermediate
coupling'' problem essentially exactly. Although the existence of
short range, spin gapped, antiferromagnetic correlations in the
half-filled two-leg Hubbard ladder and the power-law $d$-wave-like
pairing correlations in the doped ladder were initially
unexpected~\cite{DagottoScalapino}, this behavior is now well
understood~\cite{DagottoRice,Noack,BalentsFisher}. Thus it is not
surprising that a two-leg ladder model of the Fe-superconductors
exhibits short range antiferromagnetic correlations and power-law
pairing. However, it is interesting to see how such a model
captures particular aspects of the physical properties of the Fe
superconducting materials~\cite{Mazin}, specifically their spin
and pairing correlations.

We find that, in the limit of zero temperature, this model has a
diverging superconducting susceptibility with a pair structure of
a form which is the quasi-one dimensional version of the
structures which have been proposed on the basis of both weak
\cite{Mazin2,Raghu,Kuroki,Wang,Chubukov,Graser} and
strong~\cite{Hu} coupling calculations for the 2D system. We also
find that the dominant magnetic correlations are \textquotedblleft
stripe-like\textquotedblright\, with wave vector $(0,\pi )$,
reminiscent of the magnetic structure that is seen in the undoped
parent compounds (and sometimes coexisting with superconductivity)
in the Fe-pnictide superconductors. In addition, we show that the
Fe-pnictide ladder is a \textquotedblleft
twisted\textquotedblright\ version of the two-leg Hubbard ladder.
Upon untwisting the ladder, a correspondence is found with the
familiar results from cuprate related studies: the
superconductivity in the Fe ladder corresponds to
\textquotedblleft $d$-wave-like\textquotedblright\ correlations in
the cuprate Hubbard ladder, and the stripe-like magnetic
correlations are transformed into antiferromagnetic correlations
with wave vector $(\pi ,\pi )$. This supports the idea that there
is a single, unified mechanism at work in these two families of
HTSs.

\begin{figure}[t]
\includegraphics[width=0.5\textwidth]{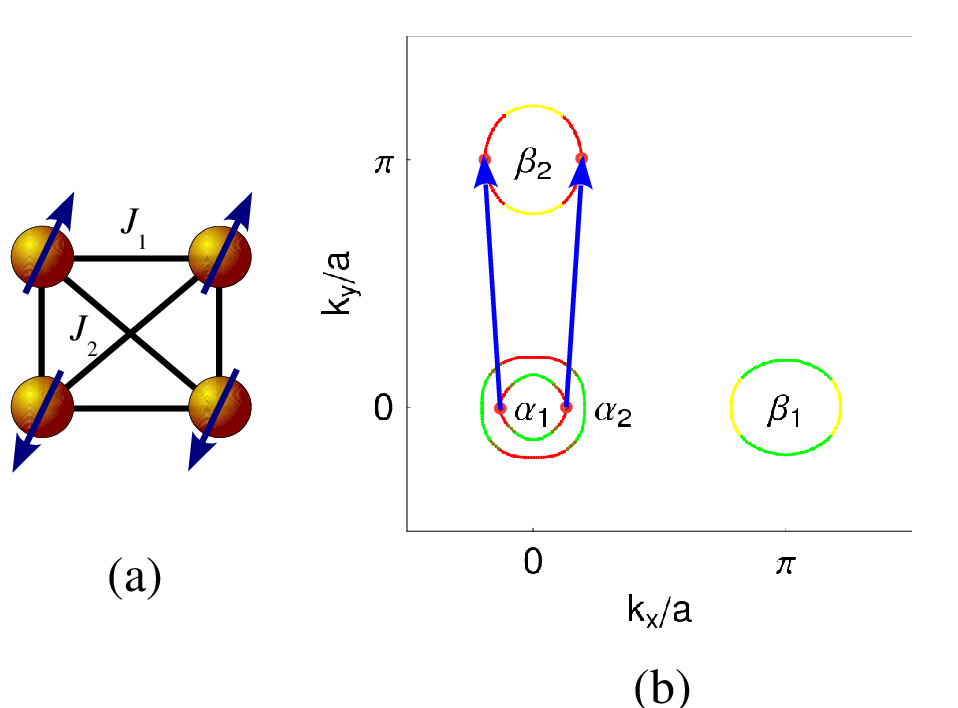}
\caption{(a) Heisenberg model showing local Fe spins coupled by
comparable nearest neighbor $J_1$ and next nearest neighbor $J_2$
exchange couplings. (b) 5-orbital tight binding Fermi surfaces
with the main orbital
contributions shown by the colors: $d_{xz}$ (red), $d_{yz}$ (green), and $%
d_{xy}$ (yellow). The arrows illustrate the type of scattering
processes which give rise to pairing in the fluctuation exchange
calculations. } \label{fig:1}
\end{figure}

As in the case of the cuprates, both strong coupling and weak
coupling models have been proposed to account for the magnetic,
structural and superconducting properties of the Fe-pnictides.
From a strong coupling perspective, the observed magnetic and
structural phase transitions and much of the dynamical magnetic
structure observed even in the superconducting phase are thought
of as arising from frustrated quantum
magnetism\thinspace\cite{Fang,Xu}. Here, as shown in Fig.
\ref{fig:1}a, one has for the undoped system a Heisenberg model in
which the second-neighbor antiferromagnetic interaction, $J_2$, is
comparable to or larger than the nearest-neighbor exchange
interaction, $J_1$. For the doped system, one has a $t-J_1-J_2$
model. Then, as discussed in Ref.~\cite{Hu}, a mean-field analysis
shows that the dominant pairing channel involves intra-orbital
$(d_{xz,\uparrow},d_{xz,\downarrow})$ and
$(d_{yz,\uparrow},d_{yz,\downarrow})$ pairs, combined to form an
$A_{1g}$ superposition. A similar pairing structure was found in
Ref. \cite{Moreo} in the large $U$ limit of an exact
diagonalization study of a $\sqrt{8}\times\sqrt{8}$ cluster. While
both of these studies only took into account the $d_{xz}$ and
$d_{yz}$ orbitals, their finding concerning the importance of
intra-orbital pairing is relevant to the model we will discuss.

Alternatively, the weak coupling picture begins by considering the
energy bands and the Fermi surface. Fig. \ref{fig:1}b shows the
Fermi surfaces for a five-orbital tight binding fit~\cite{Graser}
of a density function theory (DFT) calculation~\cite{DFT},
neglecting the effects of dispersion in the third direction,
perpendicular to the Fe-pnictide planes. Here the color indicates
the relative weights of the 3d ($d_{xz}$, $d_{yz}$, $d_{xy}$)
orbitals which form the main contribution to the Bloch
wavefunctions on the Fermi surfaces. The $d_{xz}$ and $d_{yz}$
orbitals provide the dominant weight on the hole Fermi surfaces
$\alpha_1$ and $\alpha_2$ around the $\Gamma$ point, while on the
$\beta_1$
and $\beta_2$ electron Fermi surfaces one has ($d_{yz}$, $d_{xy}$) and ($%
d_{xz}$, $d_{xy}$) contributions, respectively.

From a weak coupling perspective, it is the nesting properties of the $%
\alpha $ and $\beta$ Fermi surfaces that give rise to
the spin density wave (SDW) instability which accounts for the
magnetism of the undoped parent compounds. For the doped system,
fluctuation exchange\thinspace\cite{Raghu,Graser,Kuroki,Ikeda,Ono}
and renormalization group\thinspace\cite{Chubukov,Wang}
calculations find that magnetic fluctuations near $(\pi,0)$ and
$(0,\pi)$ lead to a pairing instability. In this case, similar to
the strong-coupling model, the
pairing arises from the type of scattering processes illustrated in Fig.~\ref%
{fig:1}b. Here, a $(\mathbf{k}\uparrow,-\mathbf{k}\downarrow)$ pair on the $%
d_{xz}$ region of the $\alpha_1$ Fermi surface is scattered by a
$\mathbf{Q}=(0,\pi)$
spin fluctuation to a $(\mathbf{Q}+\mathbf{k}^{\prime}\uparrow,\mathbf{Q}-%
\mathbf{k}^{\prime}\downarrow)$ pair on the $d_{xz}$ part of the
$\beta_2$ Fermi surface. Similar $(\pi,0)$ processes involving
$d_{yz}$ pair scattering occur between $\alpha_1$ and $\beta_1$.

\begin{figure}[t]
\includegraphics[width=0.5\textwidth]{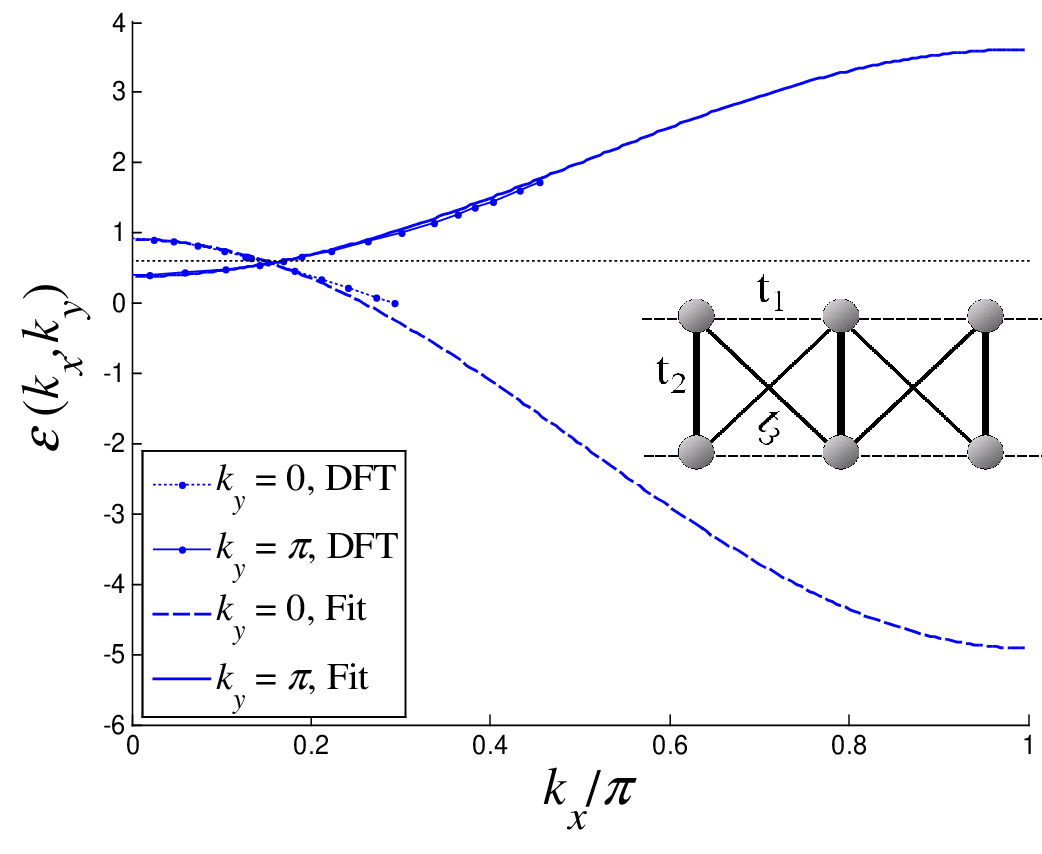}
\caption{Band structure $\protect\varepsilon(k_x,k_y)$ of the ladder with $%
t_1=-0.32$, $t_2=1$ and $t_3=-0.57$. The solid and dashed curves
correspond to $ k_y=0$ and $\protect\pi$, respectively. The black
dotted line corresponds to the Fermi energy for a filling of one
electron per $d_{xz}$ orbital. The dots represent the bands from a
DFT calculation \cite{DFT} for the 2D lattice, cut through $k_y=0$
and $\pi$. The
inset shows the ladder with the one electron hopping matrix elements $t_1$,$%
t_2$ and $t_3 $.}
\label{fig:2}
\end{figure}

\begin{figure}[t]
\begin{center}
\hskip 0.5cm \subfigure[]{\includegraphics[width=0.4
\textwidth]{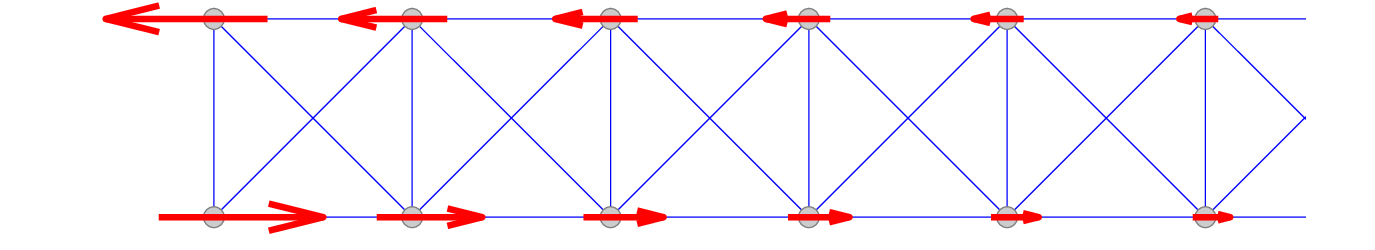}}
\subfigure[]{\includegraphics[width=0.45
\textwidth]{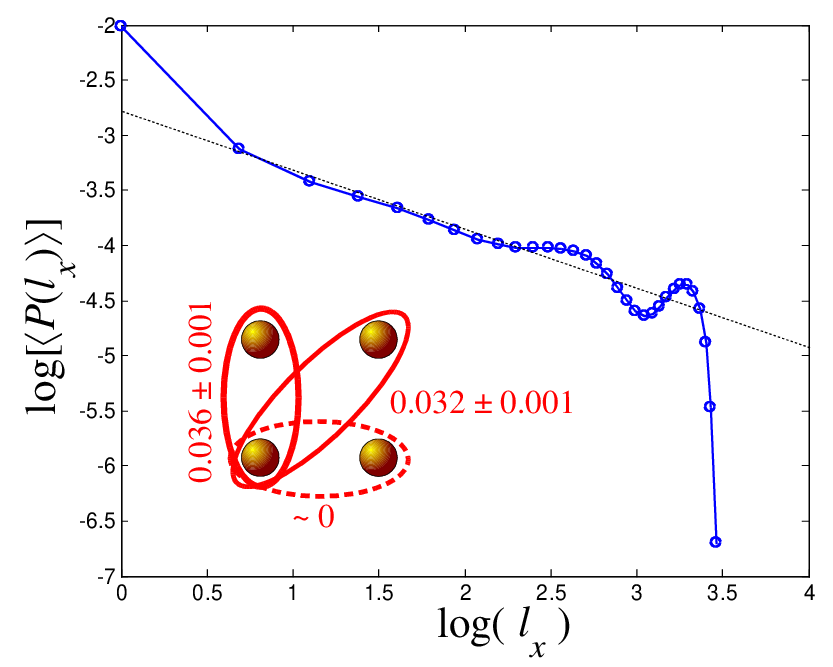}}
\end{center}
\caption{(a) Spin structure for the undoped ladder, for
$U/t_{2}=3$. The lengths of the arrows indicate the measured
values of $\langle S^z(l_x) \rangle$ in a calculation in which an
external magnetic field has been applied to the lower leftmost
site. $\langle S^z(l_x) \rangle$ decays exponentially with a
correlation length of about four sites. (b) Pairfield structure
for the $ \langle n\rangle =0.94$ doped ladder. Here, a pairfield
boundary condition has been applied to a rung on one end of the
ladder and the induced rung pairfield $\langle P_{l_x}\rangle$
versus $l_{x}$ is shown on a log-log plot. The amplitude of the
singlet pairfield across a rung and along a diagonal at $l_{x}=10$
are shown in the inset. The error bars in the pair fields were
estimated from the extrapolation of the DMRG truncation error to
zero. The pairfield amplitude along a leg is less than $10^{-3}$.}
\label{fig:3}
\end{figure}
Thus, both strong-coupling and weak-coupling approaches have been
used to describe these materials. However, what is needed is an
approach which allows one to treat the intermediate coupling
regime. Here, using DMRG, we address this issue for a caricature
of the original problem which focuses on the $d_{xz}$ orbital
$\alpha_1-\beta_2$ scattering process for $k_y=0$ and $k_y=\pi$
states near the Fermi surface. These scattering processes can be described by the two-leg ladder shown in the inset of Fig.~\ref%
{fig:2} with a Hamiltonian given by
\begin{eqnarray}
&&H=-t_{1}\sum_{i\alpha \sigma }d_{i,\alpha \sigma }^{\dagger
}d_{i+1,\alpha \sigma }-2t_{2}\sum_{i\sigma }d_{i,1\sigma
}^{\dagger }d_{i,2\sigma }
\nonumber \\
&&-2t_{3}\sum_{i\sigma }d_{i,1\sigma }^{\dagger }d_{i+1,2\sigma }+\mathrm{h.c.%
}+U\sum_{i,\alpha \sigma }n_{i,\alpha \uparrow }n_{i,\alpha
\downarrow }\text{.} \label{eq:1}
\end{eqnarray}%
Here, $\alpha =1,2$ is the leg index, $\sigma =\uparrow
,\downarrow $ is the spin index, there are leg $t_{1}$, rung
$t_{2}$ and diagonal $t_{3}$ one-electron hopping matrix elements
and an on-site Coulomb interaction $U$. The factors of $2$ in
front of $t_2$ and $t_3$ take into account the periodic boundary
conditions transverse to the ladder. The hopping strengths were
fitted to the DFT band dispersion of the $\alpha_1$ and $\beta_2$
pockets calculated in Ref. \cite{DFT}, cut through $k_y=0$ and
$k_y=\pi$ near the Fermi surfaces. At these points in momentum
space, the Bloch wavefunctions have a $ d_{xz}$ character. The
parameters we used were $t_1=-0.32$ and $ t_3=-0.57$, measured in
energy units in which $t_2=1$. The ladder bandstructure is shown
in Fig. \ref{fig:2}, where it is compared to the DFT dispersions
near the Fermi energy. The dotted line marks the location of the
chemical potential for the half-filled, one electron per $d_{xz}$
orbital case. To capture the intermediate coupling aspect of the
physics, we will take the onsite Coulomb interaction $U=3$. This
value of the interaction is smaller than the overall bandwidth of
the ladder, but larger than the Fermi energies of the hole and
electron pockets relative to their values at $k_x=0$.

DMRG calculations were carried out for a $32\times 2$ system. As
expected, there is a spin gap, and the doped system exhibits
power-law pairing correlations. At half-filling, in the presence
of an externally applied magnetic field on the first site of the
lower leg, we find the striped spin pattern shown in
Fig.~\ref{fig:3}a. The spin correlations decay
exponentially with a correlation length of about four sites and a spin gap $%
\Delta _{s}=0.14$.

For the lightly hole doped case, $\langle n\rangle =0.94$ (4
holes), we find a spin gap $\Delta _{s}=0.07$ and power-law
pairfield correlations. Here, a pairfield boundary term
\begin{equation}
H_{1}=\Delta _{1}\left( P_{1}^{\dagger }+\mathrm{h.c.}\right) \text{,}
\label{H1}
\end{equation}
with $\Delta _{1}=0.5$ and
\begin{equation}
P_{1}^{\dagger }=\left( d_{1,1\uparrow }^{\dagger }d_{1,2\downarrow
}^{\dagger }-d_{1,1\downarrow }^{\dagger }d_{1,2\uparrow }^{\dagger }\right)
\end{equation}
was added to the Hamiltonian. This acts as a proximity coupling to a rung on
the end of the ladder.

The two leg ladder is in a phase with a single gapless charge
mode, and the long range correlations are power laws characterized
by the Luttinger parameter $K_c$. For $K_c>\frac{1}{4}$ the
superconducting susceptibility is divergent, and for
$K_c>\frac{1}{2}$ it is dominant over the charge density wave
susceptibility. In addition, if $K_c>\frac{1}{2}$, the boundary
pairing term of Eq. (\ref{H1}) is relevant under a renormalization
group flow\thinspace\cite{Affleck,Feiguin} and the effective
boundary condition is described by perfect Andreev reflection. In
this case, the induced pair field decays as $\left\vert L\tan
\left( \frac{\pi l_{x}}{2L}\right) \right\vert^{-\frac{1}{4K_c}}$,
where $L$ is the length of the system\thinspace\cite{notes}. For
the stated parameters, the induced expectation value of the rung
pair amplitude $\langle P^{\dagger}(l_x)\rangle=\left\langle
d_{l_x,1\uparrow }^{\dagger }d_{l_x,2\downarrow }^{\dagger
}-d_{l_x,1\downarrow }^{\dagger }d_{l_x,2\uparrow }^{\dagger
}\right\rangle$ was measured throughout the system, and is shown
in Fig. \ref{fig:3}b on a logarithmic scale. The slope
of the curve $\log(\langle P^\dagger(l_x)\rangle)$ vs. $\log(l_x)$ gives $%
K_c\simeq 0.5$, so the ladder is near the border of the superconducting ($K_c>0.5$%
) phase. This conclusion is supported by calculations with longer
systems (up to $64\times 2 $). For electron doping, we found
similar results with slightly smaller values of $K_c$.

The inset of Fig. \ref{fig:3}b shows the amplitude for removing a
singlet pair from two sites at a position 10 sites away from the
boundary. In order to interpret these results, we perform a BCS
mean field treatment of the ladder. In such a treatment, the
amplitude for removing a singlet pair from two sites separated by
$\left( l_{x},l_{y}\right) $ is
\begin{eqnarray}
A\left( l_{x},l_{y}\right)  &\equiv &\left\langle d_{i+l_{x},j+l_{y}\uparrow
}d_{i,j\downarrow }-d_{i+l_{x},j+l_{y}\downarrow }d_{i,j\uparrow
}\right\rangle   \nonumber \\
&=&\frac{1}{N}\sum_{\mathbf{k}}{}^{^{\prime }}\frac{\Delta \left( \mathbf{k}%
\right) }{E\left( \mathbf{k}\right) }e^{i\mathbf{k}\cdot
\mathbf{l}}\text{.} \label{A}
\end{eqnarray}%
Here $E\left( \mathbf{k}\right) =\sqrt{\left[ \varepsilon \left( \mathbf{k}%
\right) -\mu \right] ^{2}+|\Delta \left( \mathbf{k}\right)
|^{2}}$, $N$ is the number of sites, and the prime on the
$\mathbf{k}$ sum implies that it is cut off when $\left\vert
\varepsilon (\mathbf{k})-\mu \right\vert >\omega _{0}$, where
$\omega _{0}$ is a cutoff energy. We assume that the gap function
has a simple form in which $\Delta (\mathbf{k})$ is approximately
constant for $\mathbf{k}$ in the neighborhood of a given Fermi
point, but with a sign change between the $k_{y}=0$ and $\pi $
bands. A reasonable estimate of the relative magnitudes of the
gaps on the two bands is obtained by requiring that, due to the
strong onsite Coulomb interaction, the onsite pair amplitude
vanishes:
\begin{equation}
A\left( 0,0\right) \simeq 2\sum_{k_{y}=0,\pi }N\left(
0,k_{y}\right) \Delta \left( k_{y}\right) \ln \left[ \frac{2\omega
_{0}}{\Delta \left( k_{y}\right) }\right] =0\text{.}  \label{A0}
\end{equation}%
Here $N\left( 0,k_{y}\right) $ and $\Delta \left( k_{y}\right) $
are the density of states at the Fermi energy and the gap
function, respectively, of the $k_{y}=0$ and $\pi$ bands. We have
assumed that $\Delta \left( k_{y}\right)
\ll \omega _{0}\ll W$ where $W$ is the bandwidth. From  Eqs. (\ref{A}) and (%
\ref{A0}), it follows then  that the ratio of the amplitude for
removing a singlet pair from a diagonal bond to removing it from a
rung, $A\left( 1,1\right) /A\left( 0,1\right) $, is approximately
$\frac{1}{2}\left( \cos k_{F}\left( 0\right) +\cos k_{F}\left( \pi
\right) \right) $\thinspace $\approx $\thinspace
 $0.9$. Here $%
k_{F}\left( 0\right) $ and $k_{F}\left( \pi \right) $ are the $k_{x}$ Fermi
momenta for $k_{y}=0$ and $\pi $, respectively. Similarly, the ratio of the
leg to rung amplitude $A\left( 1,0\right) /A\left( 0,1\right) $ is approximately $%
\frac{1}{2}\left( \cos k_{F}\left( 0\right) -\cos k_{F}\left( \pi
\right) \right) $\thinspace $\approx $\thinspace $-0.02$. These
amplitude ratios are in overall agreement with the numerical results shown in the inset of Fig. \ref%
{fig:3}b. [$A\left( 1,0\right) $ is less than $10^{-3}$, which is
essentially zero within our numerical accuracy.] The magnitude of
the pairing amplitude depends on the strength of the boundary
proximity pair field $\Delta _{1}$. Note that the rung and
diagonal pairfields are in phase with each other. A similar
pairfield pattern was found in weak coupling spin-fluctuation
calculations\thinspace \cite{Graser}. In this case, there were also $d_{yz}$%
--$d_{yz}$ orbital pairfield correlations, rotated by $90^{\circ
}$. The relative phase of the $d_{xz}-d_{xz}$ and $d_{yz}-d_{yz}$
pairfields determines whether the order parameter has $A_{1g}$ or
$B_{1g}$ symmetry. The present study can not address this issue
since it lacks the $d_{yz}$ orbitals~\cite{comment-symmetry}.

\begin{figure}[t]
\includegraphics[width=0.5\textwidth]{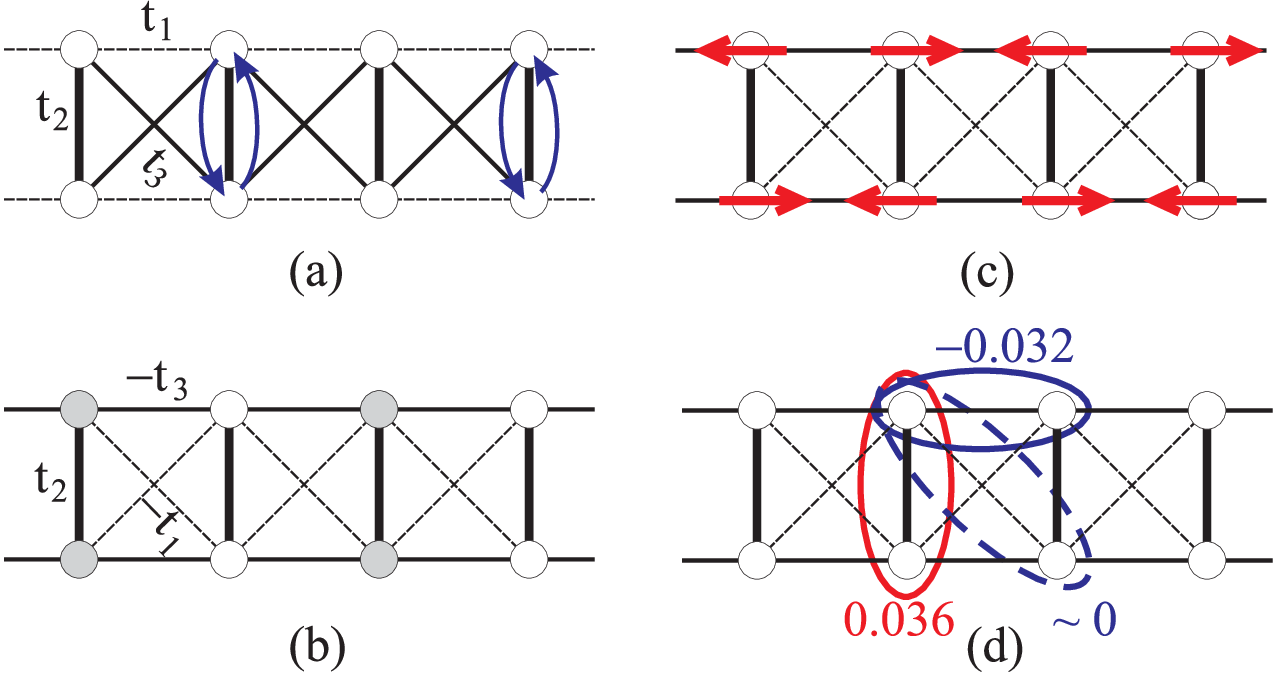}
\caption{(a-b) Mapping of the Fe Hubbard ladder, Eq. (1) to the
Hubbard ladder used to model the cuprates. (a) Twist every other
rung as indicated.
(b) Change the sign of the $d_{xz}$ orbital on the shaded sites. (c) The $(0,%
\protect\pi)$ short range spin correlations of the FeAs ladder (shown in
Fig. \protect\ref{fig:3}a) become the familiar $(\protect\pi,\protect\pi)$
spin correlations of the usual Hubbard ladder after the twists and orbital
sign changes shown in (a) and (b). (d) The pairing correlations shown in the
inset of Fig. 3b are mapped to the familiar cuprate ladder $d_{x^2-y^2}$%
--like pairfield.}
\label{fig:4}
\end{figure}

 At first sight the striped $(0,\pi)$ spin
structure of the undoped ladder and the pairing amplitudes of the
doped ladder appear different from the $(\pi,\pi)$ spin and
$d_{x^2-y^2}$--like correlations familiar for the ladders used to
model the cuprates. However, a closer look at the present model
reveals that it is in fact identical to that used to model the
cuprates.

The mapping between the Fe model we have been discussing and the
``usual" two-leg Hubbard ladder is illustrated in
Fig.~\ref{fig:4}. First, every other rung is twisted (Fig.
\ref{fig:4}a), interchanging the two sites at the ends of the rung
and, leading to a ladder having a hopping $t_3$ along the legs and
the weaker hopping $t_1$ along the diagonals. Then, as illustrated
in Fig. \ref{fig:4}b, the phases of the orbitals on the shaded
sites are changed by $-1$. This leads to the
usual Hubbard model with $t_{\mathrm{leg}}=-2t_3=1.14$, $t_{\mathrm{rung}%
}=2t_2=2$ and a next-nearest neighbor $t^{\prime}=-t_1=0.32$.
After the twist and orbital phase changes, one finds the familiar
$(\pi,\pi)$ spin correlations shown in Fig.~\ref{fig:4}c and the
$d_{x^2-y^2}$--like structure for the pairing correlations shown
in Fig.~\ref{fig:4}d.

Finally, we note that the parameters for the FeAs ladder place it
near a region of enhanced pairing for the corresponding Hubbard
ladder case~\cite{Noack}. There, it was found that the pairing
correlations are enhanced when the parameters are such that the
Fermi level is located near the top of the bonding band and the
bottom of the anti-bonding band
($t_{\mathrm{rung}}/t_{\mathrm{leg}}\lesssim2$). The proximity to
this point also makes the pairing correlations sensitive to the
bandstructure parameters. We have found that while using different
bandstructure parameters and a different $U$ does not change the
behavior found here qualitatively (e.g., the existence of a spin
gap and the pair structure is robust), the strength of the
long-range pairing correlations is very sensitive. For instance,
for the bandstructure parameters used in the present paper, the
pairing correlations are reduced upon increasing the interaction
to $U=4$. For this case, we get $K_c\approx 0.3$. In addition, if
we use the band parameters $t_1=-0.77$, $t_2=1$, $t_3=-0.65$,
obtained from a fit to a 2-orbital model~\cite{Raghu}, and
$U=4.6$, we get a significant enhancement of the pairing
correlations, and $K_c\approx 0.75$. The sensitivity of the
pairing correlations to the band parameters and to $U$ may be an
artifact of the one dimensional nature of the
ladder~\cite{comment-1d}. We believe, however, that the basic
superconducting mechanism in the ladder is operative in the FeAs
superconductors.

In summary, this analysis began by constructing a ladder model for
the Fe-pnictides with hopping parameters chosen to fit DFT
calculations. Following the results of the RPA spin-fluctuation
calculations and the mean-field $t-J_1-J_2$ and exact
diagonalization studies which indicate that the dominant pairing
involves $(0,\pi)$ $d_{xz}$ to $d_{xz}$ and $(\pi,0)$ $d_{yz}$ to
$d_{yz}$ pair scattering processes, 
only the $d_{xz}$ orbital was kept. Using the DMRG method, we
found short range magnetic correlations which mimicked the
``striped'' $(0,\pi)$ magnetic order in the undoped parent system,
and pairing correlations with a structure that agreed with both
the weak coupling RPA calculations and the strong coupling
results. The Fe-ladder turns out to be simply a twisted version of
the usual cuprate Hubbard ladder, with parameters near the regime
of enhanced pairing. We believe that this is not an accident but
arises from the close connection between the basic physics
responsible for high $T_c$ superconductivity in the cuprates
and the Fe-pnictides. 

\section*{Acknowledgement}

We thank I. Affleck and S. Raghu for useful discussions. DJS
acknowledges the Center for Nanophase Materials Science, which is
sponsored at Oak Ridge National Laboratory by the Division of
Scientific User Facilities, U.S. Department of Energy and thanks
the Stanford Institute of Theoretical Physics for their
hospitality. SAK was supported by the NSF through DMR 0758356. EB
was supported by the U.S. Department of Energy under contract
DE-FG02-06ER46287 through the Geballe Laboratory of Advanced
Materials at Stanford University.

\bibliography{FeAs}

\begin{thebibliography}{28}
\bibitem[{com({\natexlab{b}})}]{comment-1d}
\bibinfo{note}{For instance, in the ladder there is a square-root singularity
  of the non-interacting density of states at the van Hove points, which is
  known to enhance superconductivity when the chemical potential is close to
  these points. This causes an increased sensitivity to the band parameters and
  to the doping.}
\end{thebibliography}

\end{document}